\documentclass[manuscript]{aastex} 
\usepackage{graphics}

\begin{document}
\pagenumbering{arabic}

\title{SPECTRAL TYPES OF FIELD AND CLUSTER O-STARS}

\author{Sidney van den Bergh}
\affil{Dominion Astrophysical Observatory, Herzberg Institute of Astrophysics, National Research Council of Canada, 5071 West Saanich Road, Victoria, BC, V9E 2E7, Canada}
\email{sidney.vandenbergh@nrc.gc.ca}

\begin{abstract}

The recent catalog of spectral types of Galactic O-type stars by Ma\'{\i}z-Apell\'{a}niz et al. has been used to study the differences between 
the frequencies of various subtypes of O-type stars in the field, in OB associations and among runaway stars. At a high level of 
statistical significance the data show that O-stars in clusters and associations have earlier types (and hence presumably larger
 masses and/or younger ages) than those that are situated in the general field. Furthermore it is found that the distribution of 
spectral subtypes among runaway O-stars is indistinguishable from that among field stars, and differs significantly from that of 
the O-type stars that are situated in clusters and associations.The difference is in the sense that runaway O-stars, on average, 
have later subtypes than do those that are still located in clusters and associations.

\end{abstract}

\keywords{galaxies: stars: early-type}

\section{INTRODUCTION}   

Recently Ma\'{\i}z-Apell\'{a}niz et al. (2004) have published a valuable catalog of all Galactic O-type stars that have accurate spectral types. Furthermore these authors assign each of these objects to either (1) the field, (2) to associations, clusters or groupings, 
and (3) the class of runaway objects (Blaauw 1961).This information allows one to study possible systematic differences between 
the distributions of different subtypes of O-stars. In particular such information should allow one to see if there are significant
differences between the masses (or ages) of O-type stars in associations and in the field. This might provide some insight into the question whether early-type field star have a distinct evolutionary history, or whether such field O-type stars simply represent the 
debris of small clusters and associations that disintegrated a few million years ago. Finally the present data might possibly throw
 some light on the nature of the formation of runaway O-type stars (Gies 1987, Hoogerwerf, de Bruijne \& de Zeeuw 2001), the existence of which was first predicted by Zwicky (1957).  

It is hoped that the present paper will make a small contribution to our understanding of the [presently rather confused] situation regarding the suspected differences between the mass spectra (luminosity functions) of early-type stars in clusters, associations and the field.  Oey, King \& Parker (2004) have argued that there is no evidence that the massive cluster and field stars in the Small Magellanic Cloud formed in a fundamentally different way.  In the Large Magellanic Cloud de Grijs et al. (2002) find that the luminosity function slope steepen with increasing cluster radius.  For both of the Magellanic Clouds Massey (2002) finds that the luminosity function outside associations is a significantly steeper than it is in associations.  A similar result is found for Galactic massive early-type stars by Kroupa \& Weidner (2003).  In other words Oey and others prefer a uniform initial mass function, whereas Massey et al.  favor a steeper IMF in the field than in clusters and associations.

\section{DISTRIBUTION OF SUBTYPES AMONG O-TYPE STARS}   

The present work is based on the homogeneous collection of spectral types of O-type stars by Ma\'{\i}z-Apell\'{a}niz et al. (2004), which supercedes a previous catalog of older lower quality data by Garmany et al. (1982).  The statistics used in the present paper are based on the data in Table 1 of Ma\'{\i}z-Apell\'{a}niz et al.  The slightly less accurate data in their Table 2, which lists O-type stars with WR companions, were excluded from the present statistics.
The field, association (including cluster and grouping), and runaway assignments were taken from column nine of 
their Table 12. A few objects that were classified as being ``more distant than association xxx'' were excluded from the statistics.   
Our Table 1 shows the distribution of 90 field, 265 association, and 24 runaway stars over the spectral subclasses of the O-type. The most striking feature of these data (see our Figure 1 and Table 1) is that the O-type stars in associations typically have earlier 
spectral types than do those in the field.  The mean spectral type of the stars in associations is O7.3, compared to O8.2 for those located in the field. A Kolmogorov-Smirnov test shows that the probability that the association and field spectral 
types were drawn from the same parent population is only 0.03\%. The observed difference is in the sense that the O-type stars in 
associations are of earlier type (and hence younger and/or more massive) than those in the field. Such a trend would be expected if the 
field population mainly consists of disintegrated clusters and associations. This is consistent with the conclusion of Oey, King \& Parker (2004) who recently found that the most massive field stars in the Small Magellanic Cloud did not originate via a different mode of star formation 
than do the massive SMC stars in associations. In other words these authors found that it seems likely that massive field and cluster stars in both the Galaxy and 
the SMC were produced by the same star-forming process.   

The data in Table 1 can also be used to compare the frequency distribution of the 
spectral subtypes of O-type stars in the Galactic field and among runaway stars. A Kolmogorov-Smirnov test shows no statistically significant 
difference between the distributions of subtypes among field stars and among runaway stars. The statistics in Table 1 are consistent with the 
notion that runaway stars might have made a significant (but not dominant) contribution to the population of O-type stars in the Galactic field. This result is consistent with recent work by de Wit et al. (2004) who finds that the vast majority of O-type stars in the Galactic field are isolated, rather than members of associations.

Inter comparison of the distribution of subtypes among runaway O-type stars and O-type stars in associations shows that there is only a 2\% 
probability that the association members and runaway stars were drawn from the same parent population of O-type stars. Not unexpectedly the 
observed difference is in the sense that the runaway stars have systematically later spectral types, which suggests that they are on average 
either older (or less massive) than their counterparts in Galactic associations, clusters and groups. In this connection it is of interest to note that two possible runaway O2 stars just outside 
the 30 Doradus Nebula (Walborn et al. 2002) will become supernovae before they have had time to escape into the field from the 30 Dor region.

\section{CONCLUSIONS}   

From a study of the distribution of subtypes among a homogeneous sample of 379 Galactic O-type stars with accurate spectral types it is found 
that O-type stars in associations have systematically earlier spectral types than do those that populate the field. This suggests that the 
O-type stars in associations are, on average, younger (and/or) more massive than those in the field. By the same token it is found that 
runaway O-type stars typically have later spectral types, and hence are systematically older or less massive, than those that are situated in associations. The observation that massive stars in the field appear to have a steeper mass spectrum than do those in associations (Massey 1999) is consistent with the present results, which show that the O-type stars in associations are typically younger (or more massive) than those that are situated in the field.   

In summary it is found that the mass spectrum of O-type stars in associations differs from that of O-type stars in the field.  However, the present data do not allow one to say whether this effect is entirely due to evolutionary effects, or if differences in the spectra of mass formation between field and clusters also play a role.

It is a pleasure to thank Sally Oey, Nolan Walborn and a particularly helpful referee, for comments on earlier versions of this paper.

\begin{deluxetable}{lccc}
\tablewidth{0pt}    
\tablecaption{Frequency of Galactic O-stars in different environments.} 
\tablehead{\colhead{Spectral type}  &  \colhead{Field stars}  & \colhead{Associations}   &  \colhead{Runaway stars}}
\startdata

O2     &         0      &       1      &      0      \\
O2.5   &         0      &       2      &      0      \\
O3     &         0      &       7      &      0       \\
O3.5   &         0      &       6      &      0        \\
O4     &         2      &       12     &      1        \\
O4.5   &         0      &       0      &      0         \\
O5     &         2      &       18     &      1        \\
O5.5   &         0      &       2      &      0         \\
O6     &         5      &       25     &      1         \\
O6.5   &         8      &       24     &      1         \\
O7     &         4      &       35     &      0        \\
O7.5    &         7      &       14     &      3         \\
O8      &         13     &       23     &      2         \\
O8.5    &         4      &       10     &      0         \\
O9      &         13     &       35     &      5         \\
O9.5   &         19     &       35     &      7        \\
O9.7    &         13     &       16     &      3        \\
Total     &         90     &       265    &      24 
\enddata
\end{deluxetable}

\begin{figure}
\begin{center}
\resizebox{4.0in}{!}{\includegraphics{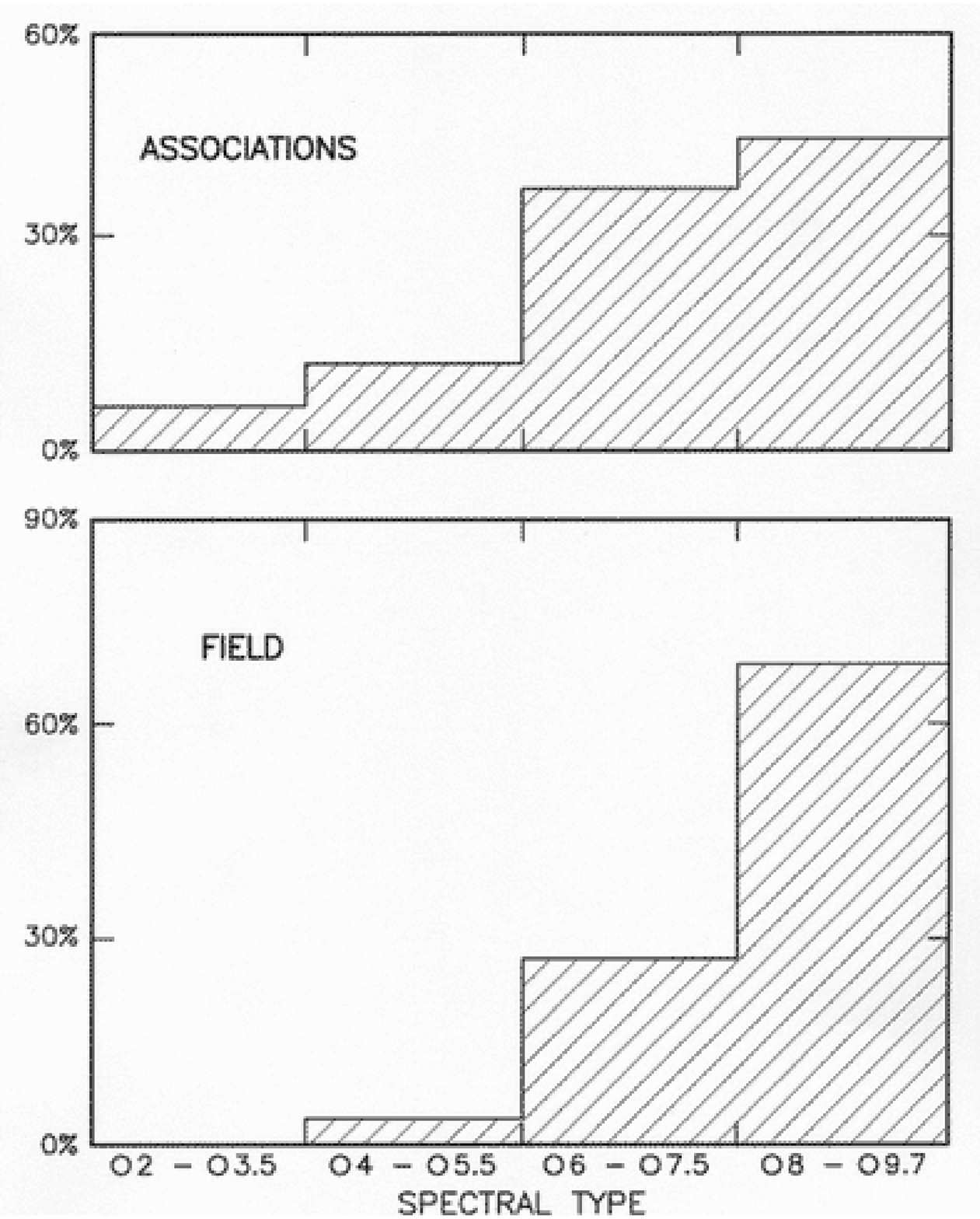}}
\end{center}

\caption{Comparison between the frequency distributions of O-type stars in associations and in the field. The figure shows that the O-stars in associations are typically of earlier spectral type, and hence younger or more massive, than those that are situated in the field.}
\end{figure}  


\begin{references} 

\reference{ }Blaauw, A. 1961, Bull. Astron. Inst. Netherlands, 15, 265
\reference{} de Grijs, R., Gilmore, G. F., Mackey, A. D., Wilkson, M.I., Beaulieu, S. F., Johnson, R. A., \& Santiago, B. X. 2002, MNRAS, 337, 597
\reference{} de Wit, W. J., Testi, L., Palla, F., Vanzi, L., \& Zinnecker, H. 2004, A\&A (in press) = astro-ph/0405348
\reference{} Garmany, C. D., Conti, P. S., \& Chiosi, C. 1982, ApJ, 263, 777
\reference {} Gies, D. R. 1987, ApJS, 64, 545
\reference{} Hoogerwerf, R., de Bruijne, J. H. J. \& de Zeeuw, P. T. 2001, A\&A, 365, 49
\reference{} Kroupa, P. \& Weidner, C. 2003, ApJ, 598, 1076
\reference{} Ma\'{\i}z-Apell\'{a}niz, J., Walborn, N. R., Galu\'{e}, H. \'{A}. \& Wei, L. H. 2004, ApJS 151, \reference{} Massey, P. 2002, ApJS, 141, 81
\reference{} Massey, P. 1999, in New Views of the Magellanic Clouds = IAU Symposium   No.190, Eds. Y-H Chu et al. (San Francisco: ASP), p.173
\reference{}Massey, P., Lang, C. C., DeGoia-Eastwood, K., \& Garmany, C. D. 1995, ApJ, 438, 188
\reference{} Oey, M. S., King, N. L. \& Parker, J. M. 2004, AJ 127, 1632
\reference{} Walborn, N. R. et al. 2002, AJ, 123, 2754
\reference{} Zwicky, F. 1957, Morphological Astronomy (Berlin: Springer) p.258

\end{references}
\end{document}